\documentclass[epj,twocolumn]{webofc}
\usepackage[varg]{txfonts}   

\usepackage{amsmath}
\usepackage{amssymb}
\usepackage{upgreek}
\usepackage[range-phrase=\times]{siunitx}
\usepackage{subfigure}
\usepackage{textcomp}
\usepackage{hyperref}

\hypersetup{
	colorlinks = true,
	breaklinks = true,
	linkcolor = black,
	menucolor = black,
	citecolor = black,
	urlcolor = black
}

\newcommand{\mueee}{$\upmu\rightarrow\text{eee}$}
\newcommand{\mueeemath}{\upmu\rightarrow\text{eee}}
\newcommand{\muposeee}{$\upmu^+\rightarrow\text{e}^+\text{e}^+\text{e}^-$}

\newcommand{\muposenunu}{$\upmu^+\,\rightarrow\,\text{e}^+\overline{\upnu}_\upmu\upnu_\text{e}$}
\newcommand{\mueeenunu}{$\upmu\,\rightarrow\,\text{e}\text{e}\text{e}\upnu\overline{\upnu}$}
\newcommand{\muposeeenunu}{$\upmu^+\,\rightarrow\,\text{e}^+\text{e}^-\text{e}^+\overline{\upnu}_\upmu\upnu_\text{e}$}
\newcommand{\muegamma}{$\upmu\,\rightarrow\,\text{e}\upgamma$}

\woctitle{Flavour changing and conserving processes}

\begin{document}

\title{Status of the Mu3e Experiment at PSI}

\author{Ann-Kathrin Perrevoort\inst{1}\fnsep\thanks{\email{perrevoort@physi.uni-heidelberg.de}}  for the Mu3e Collaboration}

\institute{Physikalisches Institut, Heidelberg University, Im Neuenheimer Feld 226, 69120 Heidelberg, Germany}

\abstract{
  The Mu3e experiment aims to search for the lepton-flavour violating decay \muposeee\ with a sensitivity to one signal decay in $\num{e16}$ muon decays at a muon stopping rate of $\SI{2e9}{\text{muons}\per\second}$. With currently available rates of $\SI{e8}{\text{muons}\per\second}$, a sensitivity on the branching ratio of $\num{e-15}$ is the aim of the first phase. This will allow for tests of new physics models with enhanced branching ratios for lepton-flavour violating processes with an unprecedented precision.\\
  The experiment must be operated at very high muon rates all the while suppressing the background of the decay \muposeeenunu\ and accidental electron-positron combinations. Therefore, a tracking detector made of thin pixel sensors with additional scintillating fibres and tiles for precise time measurement will be built. The development of the subdetectors is ongoing while detector construction is still in preparation.
}

\maketitle

\section{Introduction}
\label{intro}
\begin{figure}
	\centering
	\subfigure[The decay \mueee\ mediated by neutrino mixing in the Standard Model.]{\includegraphics[width=0.4\textwidth,clip]{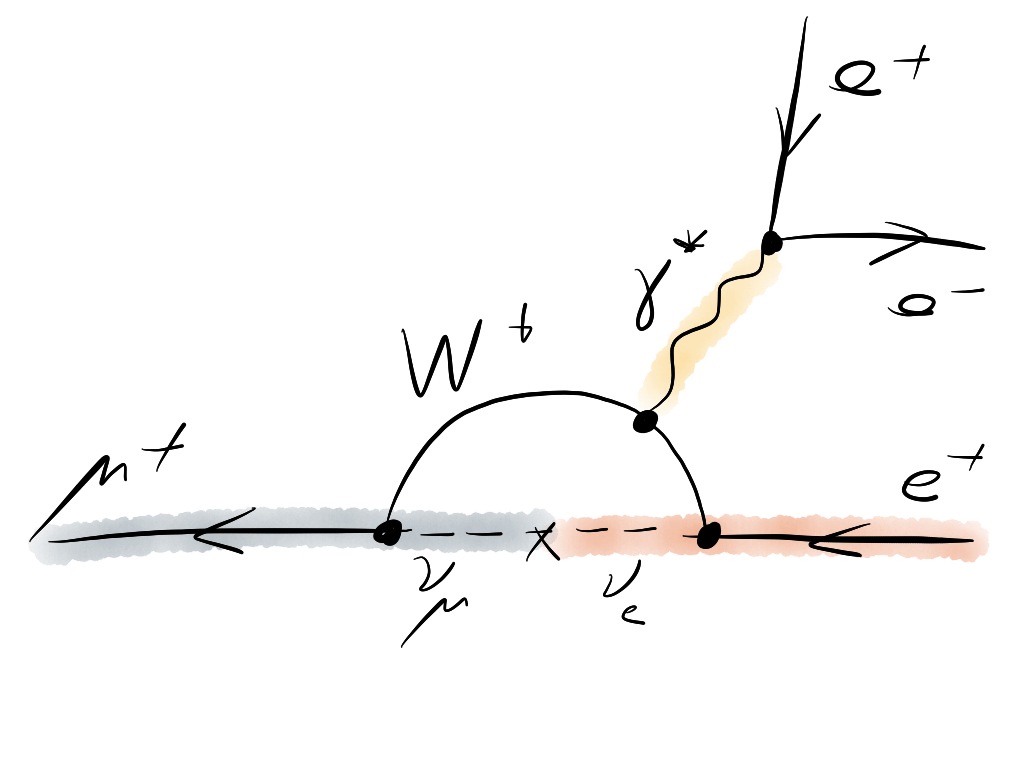}
	\label{FigMu3eSM}}\\
	\subfigure[The \mueee\ decay at loop level in BSM models.]{\includegraphics[width=0.4\textwidth,clip]{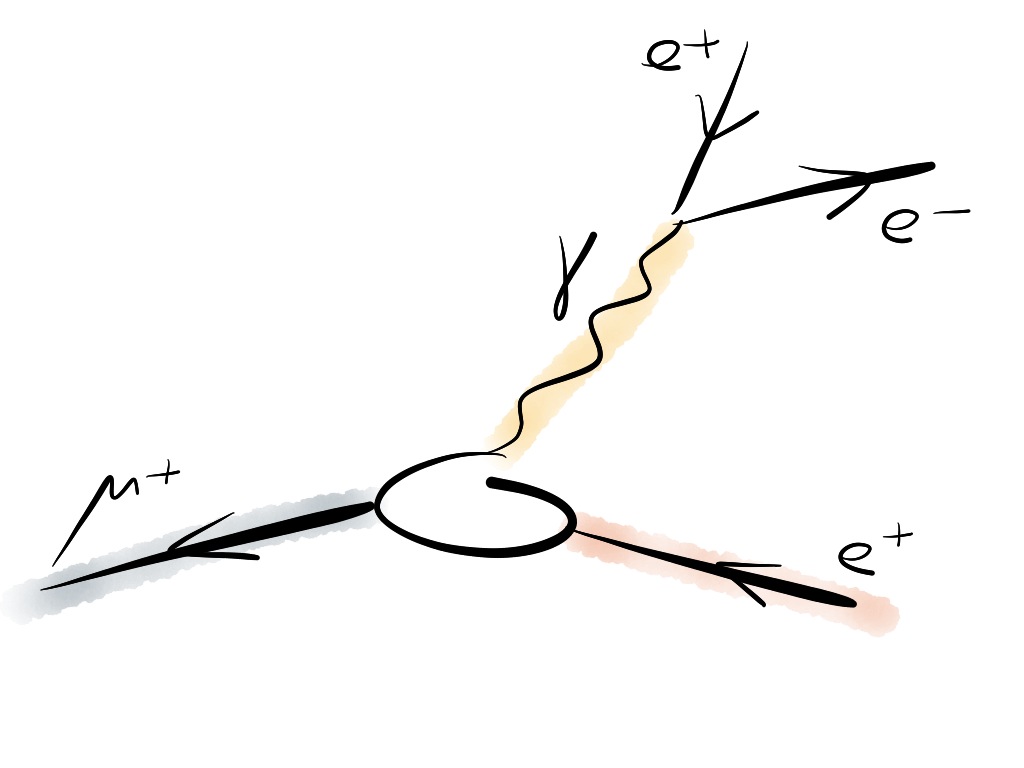}
	\label{FigMu3eLoop}}\\
	\subfigure[The \mueee\ decay at tree level in BSM models.]{\includegraphics[width=0.4\textwidth,clip]{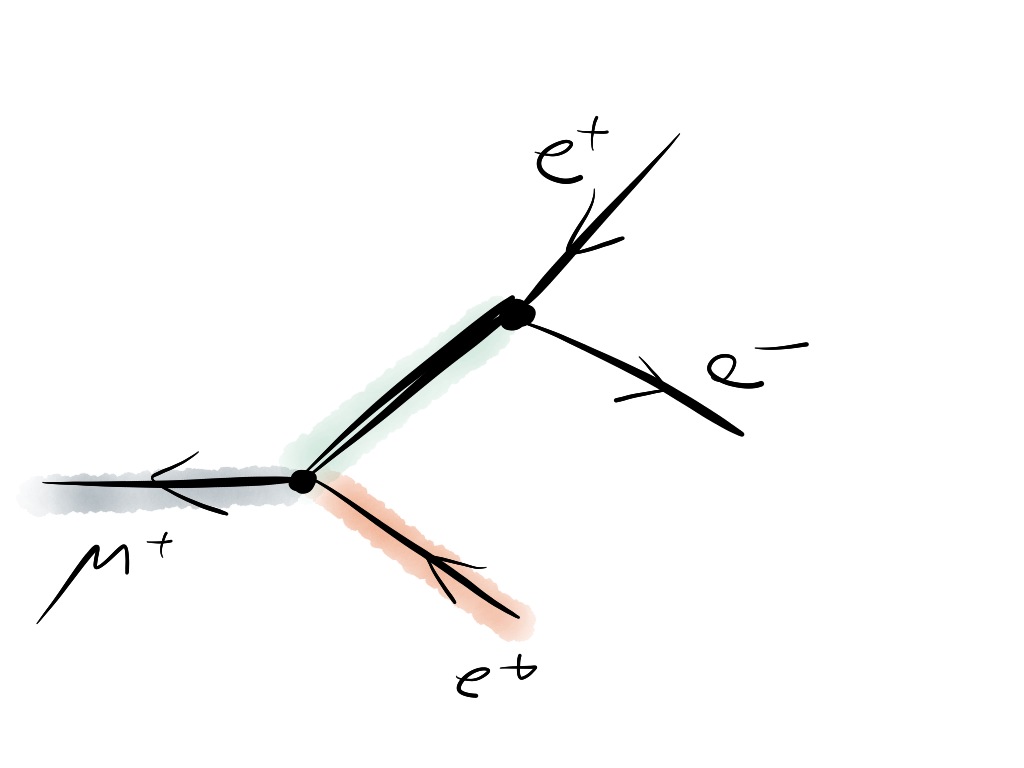}
	\label{FigMu3eTree}}
	\caption{Diagrams of the decay \mueee\ (a) mediated in the Standard Model via neutrino oscillations, and at  (b) loop and (c) tree level in BSM models.}
\end{figure}

\begin{figure}
	\centering
	\includegraphics[width=0.43\textwidth,clip]{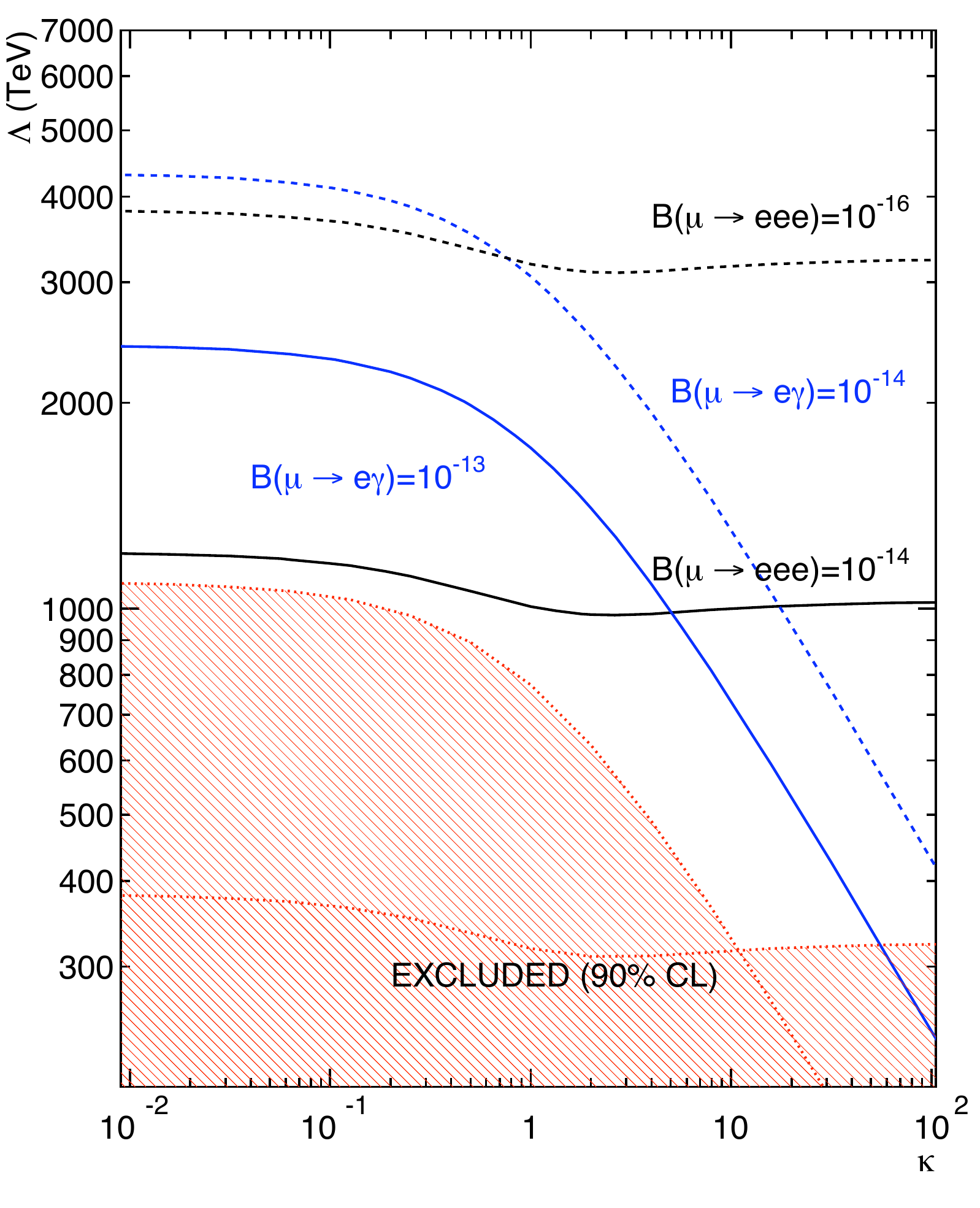}
	\caption{Comparison of the reach in effective mass scale $\Lambda$ of new physics theories in searches of the decay \muegamma\ and \mueee . Low values of $\upkappa$ represent a dominating dipole contribution whereas high $\upkappa$ values represent dominating four-fermion contact interactions~\cite{Gouvea}.}
	\label{FigGouvea}
\end{figure}

The Mu3e experiment is dedicated to search for physics beyond the Standard Model (BSM) via the lepton flavour-violating decay \muposeee . In principle, this process can be mediated via neutrino mixing~(Fig.~\ref{FigMu3eSM}), but as it is suppressed to a branching ratio below $\num{e-54}$, any observation would be a clear sign of new physics. In fact, many BSM models naturally predict enhanced branching ratios for charged lepton flavour violating processes like \mueee ~\cite{Kuno}.\\
In BSM models, the \mueee\ decay can be mediated at lowest order by dipole-like interactions~(Fig.~\ref{FigMu3eLoop}) or four-fermion contact interactions~(Fig.~\ref{FigMu3eTree}) depending on the particular model.\\
In~\cite{Gouvea}, the \mueee\ and \muegamma\ decay have been \mbox{studied} by the means of effective field theory. Depending on the underlying interaction, a search for \mueee\ with a sensitivity of $\num{e-16}$ on the branching ratio is able to test the effective mass scale of lepton-flavour violating operators in the range of several thousand $\si{\tera\electronvolt}$~(Fig.~\ref{FigGouvea}).\\
The current limit on the process \mueee\ has been set by the SINDRUM experiment to $\text{BR}_{\mueeemath}<\num{1.0e-12}$ at $\SI{90}{\percent}$ confidence level (CL)~\cite{Sindrum}. The Mu3e experiment aims to search for \mueee\ with a sensitivity on the branching ratio of $\num{e-15}$ at $\SI{90}{\percent}$ CL in the first stage of the experiment and $\num{e-16}$ at $\SI{90}{\percent}$ CL in the second stage~\cite{RP}.\\

\section{Signal and Background}
\label{signal_bg} 
\begin{figure}
	\centering
	\includegraphics[width=0.43\textwidth,clip]{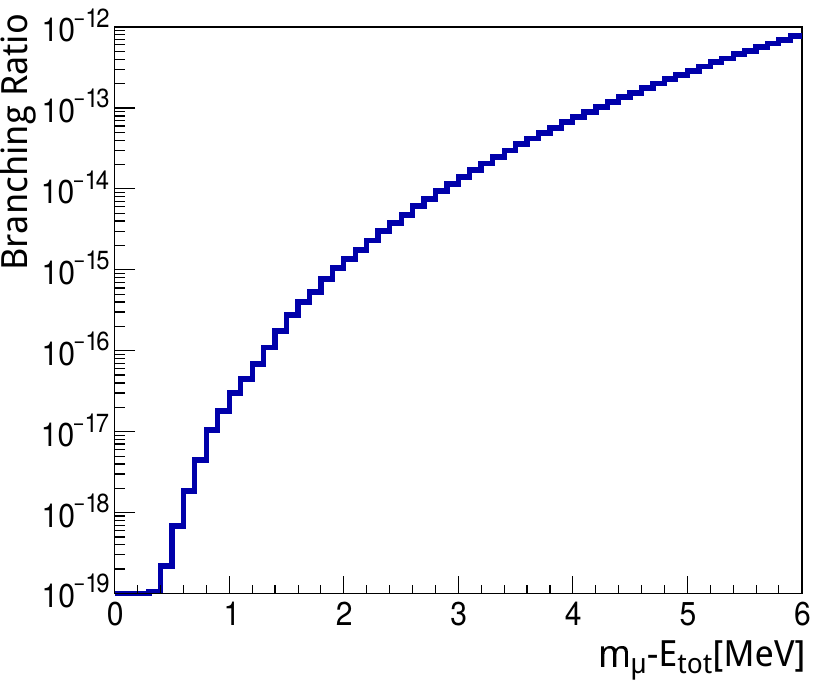}
	\caption{Integrated branching ratio for the radiative muon decay with internal conversion \mueeenunu\ for different cut values on the total energy $\text{E}_\text{tot}$ of the three decay electrons. Adapted from~\cite{Konoplich}, matrix element by~\cite{Signer}.}
	\label{FigIC}
\end{figure}

Experimentally the \muposeee\ decay is identified by measuring two positrons and an electron from a common vertex in space and time. In the Mu3e experiment the antimuons are stopped and decay at rest. Therefore, the energies of the electron and positrons sum up to the muon mass, whereas the sum of their momenta vanishes. Any other process that mimics this signature is a potential source of background and needs to be suppressed in the experiment to below the sensitivity level.\\
One source of background can arise from the accidental combination of any two positrons and an electron that within the de\-tec\-tor resolution shows the characteristics of the decay signal. As an example, this could be a positron from the dominant Michel decay \muposenunu\ in combination with a positron and electron from a Bhabha scattering event or photon conversion. The rate of the accidental combination background is dependent on the rate of incoming muons; higher antimuon rates result in more accidental combinations. This source of background can be controlled by optimized time, vertex and momentum resolutions.\\
Furthermore, the radiative decay of an antimuon with internal conversion \muposeeenunu\ has a similar signature as the decay signal. The undetected neutrinos create missing energy, thus an excellent momentum resolution is crucial in order to control this type of background. This is illustrated in Fig.~\ref{FigIC} which shows the integrated branching ratio for the internal conversion decay for different cut values on the sum of the energies of the three decay particles $\text{E}_\text{tot}$. Taking this into account, a momentum resolution of $\sim$$\SI{0.5}{\mega\electronvolt}$ is required to sufficiently suppress internal conversion decays. This is particularly challenging to achieve as the electrons\footnote{Here and in the following, electrons and positrons will simply be denoted as electrons. In the same way, muons and antimuons will be referred to as muons.}\ have a low momentum and thus are strongly affected by multiple Coulomb scattering which in turn affects the momentum measurement.\\
For the Mu3e experiment a vertex resolution of about $\SI{200}{\micro\metre}$ and a time resolution of $\sim$$\SI{100}{\pico\second}$ is envisaged.

\section{Experimental Concept}
\label{expconcept}
The detector in the Mu3e experiment has to be able to cope with high muon stopping rates while the amount of material in the active detector volume has to be kept to a minimum. With thin pixel sensors, a high momentum resolution will be achieved, while scintillating fibres and tiles will allow for precise time measurements.\\
The Mu3e detector will be built in stages corresponding to an increase in the muon stopping rate.

\subsection{Muon Beam}
\label{muonbeam}
At the Paul-Scherrer Institute (PSI) a proton beam with $\SI{590}{\mega\electronvolt}$ is provided with a very high intensity of $\SI{2.2}{\milli\ampere}$. Pions are produced on carbon targets and decay subsequently into muons. The $\uppi$E5 beam line, currently in use by the MEG experiment~\cite{MEGdetector}, is one of the highest intensity low energy muon beam lines at PSI with attainable rates of up to $\SI{e8}{\text{muons}\per\second}$ of $\SI{28}{\mega\electronvolt\per\text{c}}$ muons. It is planned that the beam time at $\uppi$E5 will be shared between the MEG and Mu3e experiments.\\
The envisaged sensitivity of $\num{e-16}$ on the branching ratio, however, can only be reached by muon rates of about $\SI{2e9}{\text{muons}\per\second}$. This would require a new beam line -- a possibility that is currently under investigation at PSI.

\subsection{Detector Design}
\label{detectordesign}
\begin{figure*}
	\centering
	\includegraphics[width=0.9\textwidth,clip]{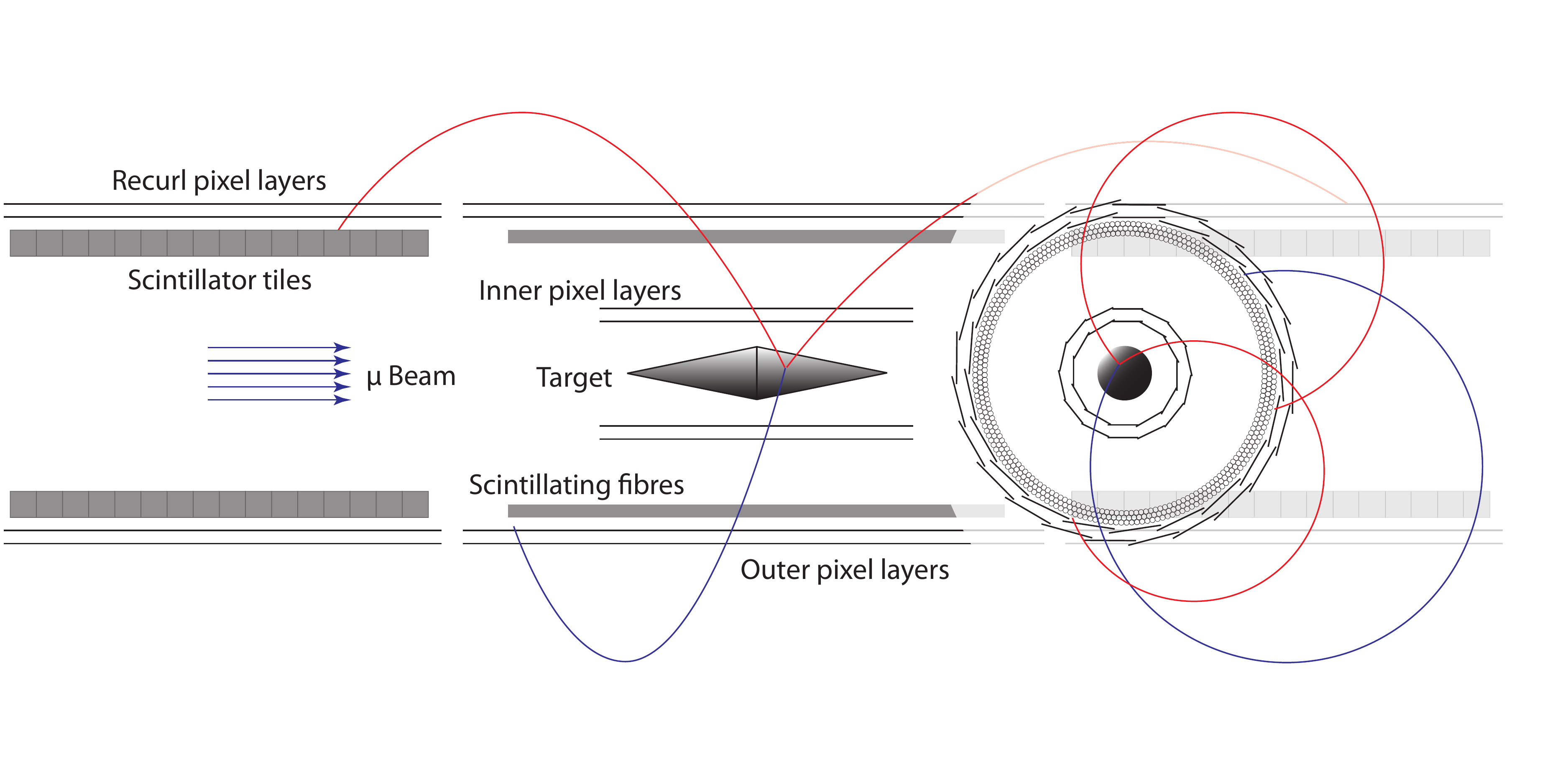}
	\caption{Schematic layout of the Mu3e detector in phase IB to be operated at $\SI{e8}{\text{muons}\per\second}$.}
	\label{FigPhaseIB} 
\end{figure*}

In the Mu3e experiment the incoming $\SI{28}{\mega\electronvolt\per\text{c}}$ muon beam will be stopped in a hollow double-cone shaped target made of Mylar. The shape of the target is chosen such that the decay vertices will spread over a large area which aids in the reduction of background from accidental combinations. As the target and detectors will be placed in a solenoidal magnetic field of $\SI{1}{\tesla}$, the momentum of the decay electrons will be measured by determining the bending radius of their trajectories. The pixel detector will be arranged in a barrel-type geometry around the beam axis~(Fig.~\ref{FigPhaseIB}).\\
In the first phase of the experiment (phase IA), there will be a double-layer of pixel sensors surrounding the target, and one double-layer at larger radii (see the central detector part in Fig.~\ref{FigPhaseIB}). The detector will be $\SI{36}{\centi\metre}$ long with a diameter of $\SI{18}{cm}$, and will initially be operated at $\SI{2e7}{\text{muons}\per\second}$. At this stage of the experiment a sensitivity to \mueee\ of $\num{e-13}$ to $\num{e-14}$ can be achieved depending on the run time.\\
In order to improve the momentum resolution the electrons will be re-measured as they move on a helical trajectory and pass again through the detector. In phase IB of the experiment, the acceptance for the re-curling tracks will be increased by installing so-called recurl-stations upstream and downstream of the central detector. Similarly to the central detector the recurl-stations will consist of a double-layer of pixel sensors and will have the same dimensions.\\
An operation at $\SI{1e8}{\text{muons}\per\second}$ will become feasible in phase IB by installing additional timing detectors: scintillating fibres in the central detector, and scintillating tiles in the additional recurl-stations. Thus, accidental combinations will be efficiently suppressed. A target sensitivity of up to $\num{e-15}$ in branching ratio is attainable in this particular detector configuration.\\
Given a new beam line, the Mu3e experiment will be operated in phase II studying $\SI{2e9}{\text{muons}\per\second}$. In order to increase the acceptance for recurling tracks, an additional pair of recurl-stations will be installed, one upstream and another downstream of the existing detector. In this configuration the Mu3e experiment can reach its target sensitivity of $\num{e-16}$.

\subsection{Pixel Sensors}
\label{hvmaps}
\begin{figure}
	\centering
	\includegraphics[width=0.45\textwidth, clip]{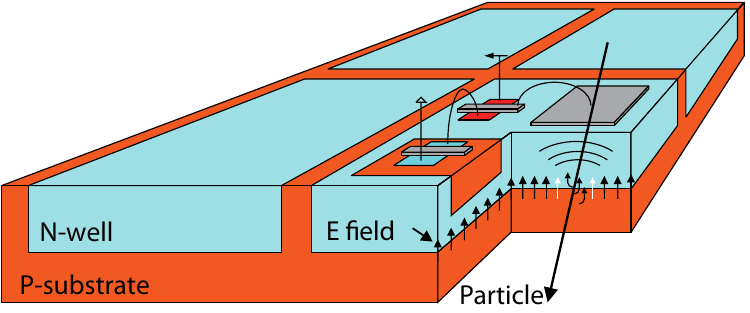}
	\caption{Schematic of a pixel sensor in HV-MAPS technology~\cite{Peric2007}.}
	\label{FigHVMAPS}
\end{figure}

The tracking detector in the Mu3e experiment will be built from silicon pixel sensors as they provide the high granularity necessary at high muon stopping rates~\cite{Berger}.\\
The High-Voltage Monolithic Active Pixel Sensors (HV-MAPS)~\cite{Peric2007,Peric2010,Peric2010b,Peric2013,Peric2015} technology has been chosen for the Mu3e experiment. The pixels are implemented as deep N-wells in a p-doped substrate and are reversely biased by about $\SI{60}{\volt}$ to $\SI{90}{\volt}$~(Fig.~\ref{FigHVMAPS}). Thus, charges generated by traversing particles are fastly collected via drift within $\mathcal{O}(\SI{1}{ns})$.\\
Moreover, the charge collection is limited to a thin depleted zone close to the surface. Therefore, the sensors can be thinned to $\SI{50}{\micro\metre}$ which allows for a very lightweight pixel detector.\\
Another advantage of the HV-MAPS is the possibility to implement analogue and digital logic directly on the sensor allowing for amplification and digitization of the signals already on the chip. No additional readout chip is needed as in standard hybrid pixel sensor designs.\\
The Mu3e experiment plans to use HV-MAPS chips of a size of $\SI[product-units = single]{2 x 2}{\centi\metre\squared}$ with a pixel size of $\SI[product-units = single]{80 x 80}{\micro\metre\squared}$ and a thickness of $\SI{50}{\micro\metre}$.\\
The latest HV-MAPS prototype for the Mu3e experiment is the MuPix7. It is a small sensor with a size of $\SI[product-units = single]{2.9 x 3.2}{\milli\metre\squared}$ and a pixel size of $\SI[product-units = single]{103 x 80}{\micro\metre\squared}$, but features already the full functionality of the final chip including an on-chip readout state machine. The output of the MuPix7 is zero-suppressed hit addresses and corresponding time stamps transmitted via a fast serial link at $\SI{1.25}{\giga\text{bit}\per\second}$.\\
The MuPix prototypes have been extensively tested in the laboratory and in testbeam campaigns~\cite{Shrestha,Wiedner2015}. For MuPix7, efficiencies above $\SI{99}{\percent}$ for the entire test system at a noise rate of less than $\SI{100}{\hertz}$ per pixel have been measured as well as a time resolution of better than $\SI{12}{\nano\second}$. Thus, already the MuPix7 prototype  meets all the requirements on the pixel sensors. 

\subsection{Lightweight Mechanics and Cooling}
\label{mechanics}
For the $\SI{50}{\micro\metre}$ thin pixel sensors $\SI{100}{\micro\metre}$ thick Kapton based flexprints with aluminum or copper traces will be utilized to connect the signals to the front-end electronics and to power the sensors. In addition, the sensors with flexprints will be glued to a mechanical support structure built of $\SI{25}{\micro\metre}$ Kapton foil. Thus, one layer of the pixel detector will contribute to the overall material budget with about $\SI{1}{\textperthousand}$ of a radiation length.\\
The sensors and electronics dissipate heat and need to be cooled. In the case of the Mu3e experiment, gaseous helium is chosen as a coolant in order to minimize multiple Coulomb scattering in the detector volume.\\
The cooling concept has been studied in simulation and tested with mechanical prototypes built of Kapton and thin glass plates~\cite{Wiedner2014}.

\subsection{Scintillating Fibres and Tiles}
\label{scintillators}
In order to improve the time resolution, the detector will be equipped with scintillating detectors from phase IB onward. In the most central part of the detector, three layers of $\SI{250}{\micro\metre}$ thick scintillating fibres will be used with silicon photo-multiplier (SiPM) readout at both ends. Fibres with round and squared cross sections are currently under study. Current prototypes of scintillating fibre modules yield a time resolution of below $\SI{1}{\nano\second}$~\cite{Bravar}.\\
In the recurl-stations the amount of material is no longer an issue and therefore scintillating tiles of about $\SI{0.2}{\centi\metre\cubed}$ will be used. The tiles will also be read out by SiPMs. Current prototypes yield a time resolution of better than $\SI{100}{\pico\second}$~\cite{EckertPhD}.\\
The readout of the SiPMs of both fibres and tiles will be performed with a custom-designed ASIC, the STiC chip~\cite{Shen, Harion}.

\subsection{Data Acquisition}
\label{daq}
In the final phase of Mu3e the detector will consist of nearly 300 million \mbox{pixels}, 4000 scintillating fibres and 7000 scintillating tiles.\\ 
The data acquisition will be performed without a hardware trigger~\cite{Wiedner2014B}. All sub-detectors will continuously send zero-suppressed data to the readout chain. At $\SI{2e9}{\text{muons}\per\second}$, this will account for a data rate of about $\SI{1}{\tera\text{bit}\per\second}$. The data will then be forwarded to a filter farm consisting of PCs equipped with graphics processing units. On the filter farm, fast online track reconstruction will be performed and signal-like topologies will be searched for. By this means, the data rate written to mass storage will be reduced to less than $\SI{100}{\mega\text{B}\per\second}$.

\section{Sensitivity Studies}
\label{sensitivity}
\begin{figure}
	\centering
	\includegraphics[width=0.475\textwidth,clip]{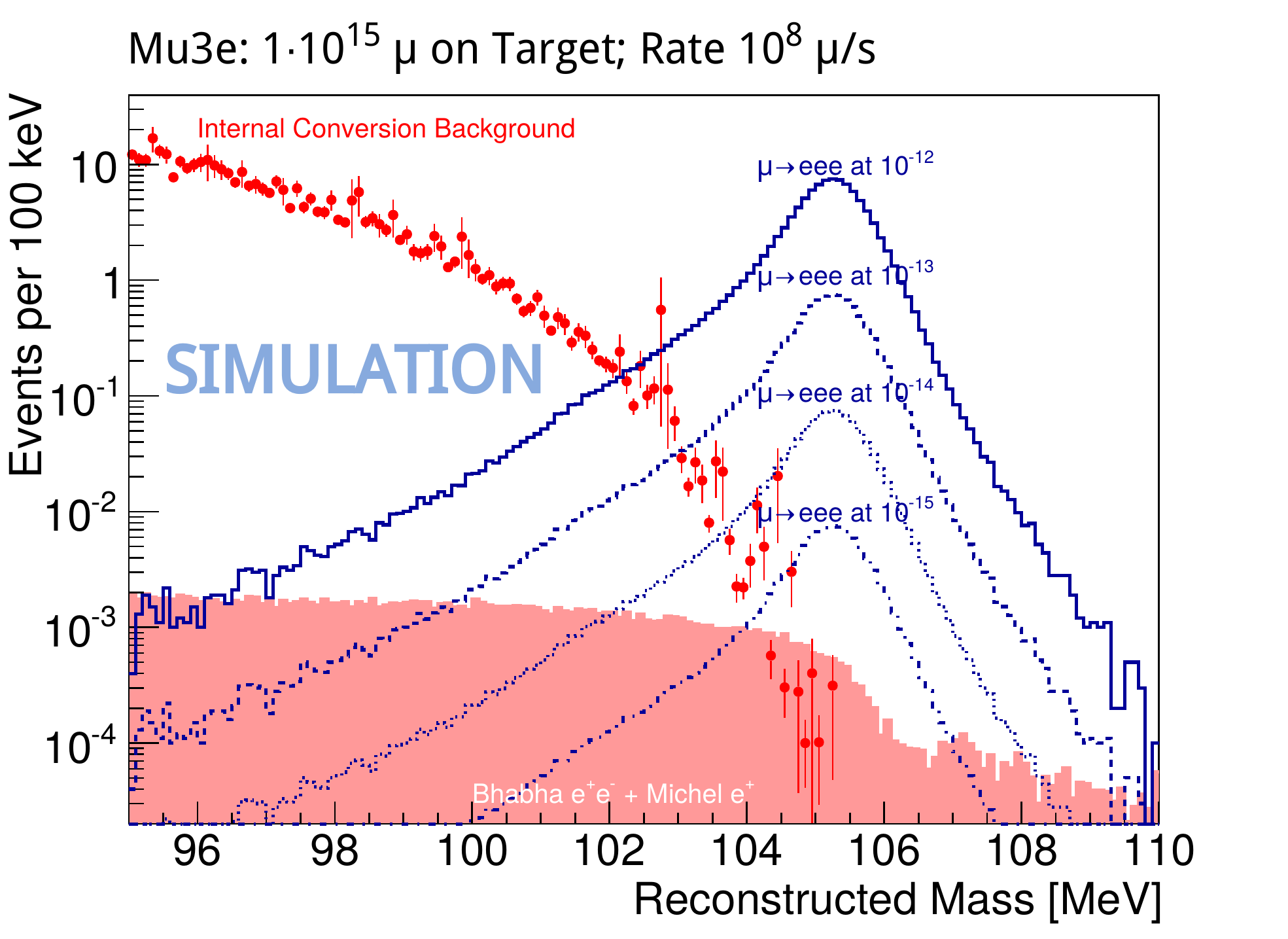}
	\caption{Reconstructed mass of three electrons from simulated internal conversion events, accidental combinations of Bhabha scattering events and Michel decays. In comparison, the \mueee\ decay at various branching ratios is shown.}
	\label{FigSensitivity}
\end{figure}
In order to estimate the detector performance, a full Geant4~\cite{Agostinelli,Allison} based simulation of the Mu3e detector has been established.\\
The result for the phase IB detector configuration with $\num{e15}$ muons on target at a rate of $\SI{e8}{\text{muons}\per\second}$ can be seen in Fig.~\ref{FigSensitivity}. It shows the reconstructed invariant mass of three electrons in the signal region from different sources of background, namely the radiative decay with internal conversion and accidental combinations of electrons from Bhabha scattering and the Michel decay, as well as signal events implemented as three body decays at different branching ratios. The detector performs well enough to separate signal decays from background even at a branching ratio of $\num{e-15}$.

\section{Status and Outlook}
\label{outlook}
The research proposal~\cite{RP} for the Mu3e experiment was approved at PSI in 2013. At the moment, the technical design report is being prepared and will be presented at PSI in 2016.\\
Research and development on the subdetectors as well as the preparation of the detector construction is ongoing. Commissioning of the detector and first data collection is expected for 2017.

\appendix
\section{Acknowledgements}
The author acknowledges support by the DFG Research Training Group on Particle Physics beyond the Standard Model.

\end{document}